\documentclass[sigconf,natbib=false]{acmart}

\AtBeginDocument{%
  \providecommand\BibTeX{{%
    Bib\TeX}}}

\setcopyright{rightsretained}
\copyrightyear{2024}
\acmYear{2024}

\RequirePackage[
  datamodel=acmdatamodel,
  style=acmnumeric,
  ]{biblatex}

\begin{document}

\title{Fostering Inclusion: A Regional Initiative Uniting Communities to Co-Design Assistive Technologies}



\author{Katharina Schmermbeck}
\orcid{0009-0005-8948-544X}
\affiliation{%
  \institution{University of Innsbruck, Department of Mechatronics, Chair of Production Technology}
  \streetaddress{Technikerstr. 13}
  \postcode{6020}
  \city{Innsbruck}
  \country{Austria}
}
\email{katharina.schmermbeck@uibk.ac.at}

\author{Oliver Ott}
\orcid{0000-0002-7680-7893}
\affiliation{%
  \institution{University of Innsbruck, Department of Mechatronics, Chair of Production Technology}
  \streetaddress{Technikerstr. 13}
  \postcode{6020}
  \city{Innsbruck}
  \country{Austria}
}
\email{oliver.ott@uibk.ac.at}

\author{Lennart Ralfs}
\orcid{0000-0003-2171-1883}
\affiliation{%
  \institution{University of Innsbruck, Department of Mechatronics, Chair of Production Technology}
  \streetaddress{Technikerstr. 13}
  \postcode{6020}
  \city{Innsbruck}
  \country{Austria}
}
\email{lennart.ralfs@uibk.ac.at}

\author{Robert Weidner}
\orcid{0000-0003-1286-4458}
\affiliation{%
  \institution{University of Innsbruck, Department of Mechatronics, Chair of Production Technology}
  \streetaddress{Technikerstr. 13}
  \postcode{6020}
  \city{Innsbruck}
  \country{Austria}
}
\affiliation{%
  \institution{Helmut Schmidt University (HSU), Faculty of Mechanical Engineering, Laboratory of Production Engineering (LaFT)}
  \streetaddress{Holstenhofweg 85}
  \postcode{22043}
  \city{Hamburg}
  \country{Germany}
}
\email{robert.weidner@uibk.ac.at}




\renewcommand{\shortauthors}{Schmermbeck et al.}

\acmConference[A3DE @ HRI '24]{Assistive Applications, Accessibility,
and Disability Ethics @ 19th Annual ACM/IEEE International Conference on Human Robot Interaction}{March 15, 2024}{Boulder, CO}

\begin{abstract}

People with disabilities often face discrimination and lack of access in all areas of society. While improving the affordability and appropriateness of assistive technologies can pave the way for easier participation and independence, awareness and acceptance of disability as part of society are inevitable. The presented regional initiative strives to tackle these problems by bringing together people with disabilities, students, researchers, and associations. During different lecture formats at the university, students co-design assistive technologies with people with disabilities. After one year in practice, we reflect on the initiative and its impact on assistive technology development and mitigation of ableism. We conducted and analyzed thirteen semi-structured interviews with participants and other involved stakeholders. Not all co-design projects were finished within the time of a lecture. Participants nevertheless appreciated the co-design approach and steps in the right direction as projects are continued in upcoming semesters. Interviewees highlighted the initiative's importance in raising awareness and broadening knowledge regarding disability and internalized ableist assumptions for those participating. We conclude that collaboration, continuity, and public outreach are most important to work towards tangible assistive technologies, bridging accessibility gaps, and fostering a more inclusive society.

  \end{abstract}



\received{05 February 2024}
\received[accepted]{07 March 2024}

\maketitle

\section{Introduction}

Because of their pivotal role in promoting the autonomy and independence of people with disabilities, providing access to appropriate assistive technologies (ATs) is essential. To avoid technology abandonment and align functionalities with individual needs, end-users must be involved in the design process \cite{phillips_predictors_1993}. Inspired by the Do-It-Yourself and makerspace community, novel approaches aim at empowering people with disabilities to become makers of their own ATs. However, due to limited access to suitable facilities, technical knowledge, and time needed, individuals with disabilities often face difficulties in initiating or executing projects independently \cite{allen_barriers_2023, higgins_towards_2023, thorsen_patient_2021}. Co-design workshops strive to overcome the problems by building interdisciplinary design teams with end-users that combine the expertise of their members \cite{aflatoony_at_2020, narain_athack_2020}. 

Open challenges of workshops are how to avoid stigmas and assumptions of able-bodied people as well as unhealthy power dynamics during the collaborative design processes \cite{gerling_reflections_2022, shew_ableism_2020, fraser-barbour_shifting_2023}. As the need for appropriate, accessible, and affordable ATs remains, the regional initiative \textit{INNklusion} strives to find ways to foster anti-ableist co-design approaches. During two lecture formats, students and people with disabilities develop solutions for various challenges. The process is continuously supported and guided by advocacy groups for people with disabilities, associations, and research experts from different fields. The initiative combines the lectures with a regular open forum, regional networking and public outreach activities to create awareness and discussion. By conducting thirteen interviews with different participating stakeholders, we analyze how the initiative's co-design lectures contribute to creating appropriate assistive technologies for individuals with disabilities. Additionally, we assess the role that the initiative can play in reducing ableism.

\begin{figure*}[h!]
      \centering
      \includegraphics[width=\linewidth]{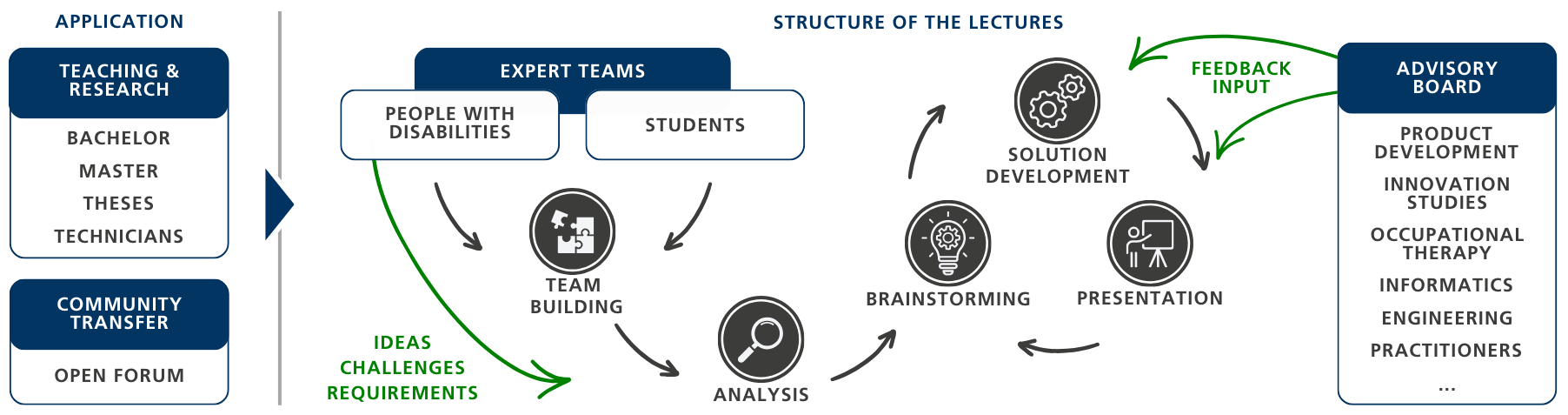}
      \caption{The regional initiative incorporates teaching and research for the development of ATs and a regular open forum to support the exchange of ideas. During the lectures, expert teams of students and people with disabilities iteratively develop ATs. Experts from various fields support the design process during the semester.}
      \label{fig:Structure_Participants}
      
      \Description{The left side of the graphic shows the parts of the initiative in blocks. The first block is "Teaching and Research", which consists of Bachelor and Master lectures, student theses, and Technicians supporting AT developments. The second block, the "Community Transfer", is conducted by a regular open forum where interested parties can join. The right side of the picture shows the structure of the lectures. Starting with the stakeholders: people with disabilities and students form the expert teams. As a team, they analyze the ideas, challenges, or requirements of people with disabilities, brainstorm and develop solutions together, that are iteratively presented. Experts from various fields form the advisory board and assist the development process through specific input and feedback. The advisory board consists of people from product development, innovation studies, occupational therapy, informatics, engineering, practitioners, and others.}
\end{figure*}

\section{Background and Related Work}
This section provides an overview of relevant trends in the development of assistive technologies as well as different co-designing approaches and their strengths and weaknesses.

\subsection{Appropriate Assistive Technologies}

The term assistive technologies refers to a broad range of physical products, software, tools, or systems that are designed to enhance and enable the participation and inclusion of individuals with disabilities or elderly in all aspects of society \cite{world_health_organization_global_2022-1, smith_assistive_2022}. Being essential for many different areas of life, ATs can improve their users' independence, autonomy, and perceived security \cite{van_dam_impact_2023}. Wheelchairs, crutches or grab rails facilitate mobility, and hearing aids or voice assistance tools can enhance the sensory experience. Small devices, such as a key turning device or a spoon stabilizer, can support activities of daily living. Besides problems in accessibility and affordability of ATs, they are often not appropriate for the specific needs of users. Reasons can be that the solutions do not function as needed or individual requirements have changed since their acquisition \cite{de_witte_assistive_2018}. Early involvement of end-users in the design process has shown to produce better solutions for the fit of individual needs and reduce the likelihood of technology abandonment \cite{bohre_review_2023}.

Additive manufacturing can offer the possibility to build customized and relatively cheap AT components from digital models built in computer-aided design (CAD) software \cite{abdi_emerging_2021,thorsen_patient_2021}. Platforms such as \textit{Thingiverse} share user-created designs, which enables a broad community to independently print products without the necessity for prior CAD designing skills \cite{buehler_sharing_2015, parry-hill_understanding_2017}. Building upon the same idea, initiatives like the Canadian \textit{Makers Making Change} \cite{neil_squire_society_makers_2024} share digital models of ATs. However, shared ATs often need to be adapted to personal requirements, which needs knowledge on how to use CAD software or is impossible due to file formats given. \textit{Makers Making Change} tries bridging this gap by connecting people with CAD skills and end-users over their platform. However, it is important to give access to information and tools to enable people with disabilities to build up skills and technical knowledge themselves \cite{allen_barriers_2023}. 




\subsection{Co-Designing Assistive Technologies}

Co-design enables users and other people from relevant fields to extensively collaborate in the design process pooling their knowledge, interests and experiences. Emerging from user-centered design approaches, where users take part as external but rather passive advising experts, co-design strives to give collaborators diverse roles based on their individual skill set and fosters strong involvement during the entire process \cite{sanders_co-creation_2008}. This includes decision making and ensuring the design outcomes are aligned with priorities and requirements of people with disabilities.

There are various approaches that apply co-design to develop ATs. Case studies that explored AT design with end users and health care professionals indicate that the collaboration led to appropriate designs and supports a combination of participants' expertise \cite{aflatoony_at_2020, howard_assessing_2022}. However, co-designing ATs within the health care sector, e.g., in rehabilitation centers, is currently confined to occasional research studies. Hackathons, such as \textit{ATHack} \cite{narain_athack_2020}, bring together people with disabilities with developers and makers from various fields to create ATs during a short period of two or three days. Often, however, people with disabilities are integrated rather as idea providers than real collaborators \cite{allen_barriers_2023}. \textit{HackaHealth} focuses their hackathons and university lecture for engineering students on exchange and joint work with people with disabilities \cite{hackahealth_hackahealth_nodate}. It has been shown that co-designing in a university setting is beneficial for participating students to learn and gain awareness about disability and accessible design \cite{higgins_towards_2023, putnam_best_2016, shinohara_tenets_2018}. Moreover, co-design approaches tend to be most effective if they include people from various disciplines, e.g., engineering, informatics, social and clinical sciences, business, and different sectors, e.g. industry, government, and research. It encourages creative and holistic thinking, ultimately leading to more innovative solutions \cite{boger_principles_2017}. 

Co-designing approaches require careful planning and guidance. Unhealthy power dynamics within teams and implicit views of able-bodied team members can easily undermine a respectful, anti-ableist, and real collaboration, where members with disabilities have the power to make decisions over the design directions \cite{shinohara_tenets_2018}. Gerling et al. recommend to proactively and continuously reflect on and investigate all participants' implicit views and attitudes \cite{gerling_reflections_2022}.

\section{The initiative's approach}

Building upon the best practices and current challenges, the presented initiative pursues a multifaceted approach. As depicted in figure \ref{fig:Structure_Participants}, it combines teaching and research on ATs with networking and public outreach activities. AT projects, that have been designed, encompass a wide range, from tangible solutions, like a hairband for one-handed use, a bed sheet covering aid, a communication tool for a locked-in syndrome person, to intengible solutions, such as implementing a \textit{Silent Hour} for low-stimulus environments in grocery shops.

The initiative organizes two lectures, a Bachelor course for engineering students and a course open to all Master students of the university. In the former mandatory course, that teaches CAD and additive manufacturing, students engage in a six-week group project to co-design small mechanical ATs with people with disabilities. During the latter course, students of various disciplines form interdisciplinary groups to work with people with disabilities on a chosen project. The project work of the two lecture formats differs in type, complexity, duration, and team composition but share the same methodology. As shown in figure \ref{fig:Structure_Participants}, the co-design approach consists of multiple stages. To actively reflect upon internalized attitudes, the lectures start by discussing ableism and barriers to inclusion with participants. This first meeting is held by the head of the university's disability office and multiple people with disabilities. To ensure rules are followed throughout the semester, each \textit{expert team} is asked to define a set of rules for their collaboration. Members are encouraged to express what is important for them when working with others and, if necessary, adapt rules over the course of their work. If conflicts occur, participants can contact a trusted third party. The project work starts with the \textit{expert team} selecting a design challenge based on their interests. In an iterative design process, they analyze requirements to build and test prototypes. During weekly meetings, \textit{expert teams} can present and reflect on their progress while being supported and guided by the lecturers and an advisory board of experts from different fields. People with disabilities voluntarily participate in the lectures. Thus, it is important to find a balance between in-person meetings and their considerable time commitment alongside work or study responsibilities. In the past semesters, it was collectively decided that participation would occur bi-weekly. Nevertheless, the teams regularly met at other locations to test prototypes and concepts. As the complexity and requirements regarding the safety and durability of the design challenges vary, the projects may result in prototypes. These then form the basis for other group developments in the upcoming semesters, or they are continued by a team of three student technicians employed by the initiative. All solutions are provided to participants free of charge.

Alongside the university lectures, the initiative holds a monthly open forum for interested people to get to know each other in a casual setting. Here, people introduce and exchange ideas, discuss topics of ATs, and initiate possible cooperation. It strives to lower barriers for people to engage during the lectures and to strengthen the initiative's public outreach. Additionally, we engage in active partnerships with associations for people with disability, the regional government and other universities to further develop and expand 
the initiative.

\section{Interviews and Data Analysis}

To evaluate the initiative after one year in practice, the first author conducted semi-structured interviews with different stakeholder groups. Thirteen people volunteered to take part in either in-person or phone interviews. The participants are four people with disabilities (PWD-1 to PWD-4), one person who works as a personal assistant representing their employer during the lecture (PA-1), five students (ST-1 to ST-5), two professionals for AT research (PAT-1, PAT-2) and the head of the university's disability office (DO-1).

The interviews, which took around thirty minutes on average, were structured in three parts. Starting with an introduction, participants were questioned about the general experience of participating in the lecture. Depending on the participants role, questions altered. For example, PWD, PA, DO, and ST got questions regarding their personal and professional learning outcomes. PATs, PWD-1, and PWD-2 were asked about the lecture's impact on their work in associations and research. Additional questions delved into the experience of working in collaborative and interdisciplinary teams. The second part of the interview concerned the personal experience and contribution to the development of ATs, the outcomes of developments, and their influence on the inclusion of people with disabilities. Lastly, interviewees were encouraged to talk about ableist structures, existing environmental and social barriers for people with disabilities, and their perspectives on addressing these challenges.

Interviews have been recorded, transcribed, and iteratively coded by the first author. During the first deductive coding cycle, the data was sorted into two categories aligned with the research questions. The second coding round used open coding and in vivo coding to find emerging ideas and topics from the data. Lastly, codes were reviewed, merged, and organized to extract patterns and themes.

\section{Findings}
In the following subsections, the main results extracted from the interviews are presented. If direct quotes are given, they have been translated from the interview language. 

\subsection{Co-Designing ATs}
\textbf{Guided co-designing creates new opportunities - }All ten interviewees who actively worked on an AT project (PA-1, PWDs, STs), positively acknowledged the co-design approach. Two of the PWDs stated that while they have previously built many ATs themselves, they were seeking additional technical knowledge to realize some of their ideas. When asked about their experience, all PWDs felt that the continuous participation in the design process helped to fulfil their requirements. PWD-4 said, "The students [...] viewed me more as an enrichment and considered me an expert." Nevertheless, a frequent guidance by the advisory board is important to ensure safe solutions. PAT-2 expressed their concern about taking people with disabilities as the only source for defining requirements as "One must also consider cognitive limitations and the fact that one's own body perception may be impaired, for example, after a stroke." PWD-2 felt they and their team would have needed advice from physio- or occupational therapists already when brainstorming about possible directions. Students ST-1 and ST-2 liked the direct exchange with the end user to iteratively and extensively test their prototypes in real scenarios as the results were decisively guiding their developments. Moreover, working as a team eased the search for possible solutions. ST-5 said, the joint work "was very helpful, because it was only through that that we came up with other ideas." Moreover, eight of ten actively designing participants (PWDs, STs, PA) felt that the personal exchange and the real challenge contributed to everyone's motivation and drive throughout the whole semester. 

\textbf{Co-design lectures do not yield products -} The outcomes of the designs were seen with mixed feelings. Especially the short amount of time for designing was mentioned by five interviewees as reason leading to the unfinished status of projects. Moreover, the professional background of group members (PWD-2, PWD-4, ST-5) and requirements for the students to document their work for grading purposes (PWD-2) were named as factors slowing down the work progress. Nevertheless, all participants were content with the status of their AT, because "one has done a step into the right direction" (PWD-4) and "everything is better than before" (PWD-2). PAT-1 added that "You can't expect a product to come out of it because product development is a completely different world." As PWDs requested, AT developments are continued until the satisfaction of the end users in following lectures or with the student technicians during their working hours.

\textbf{Open forum complements the lectures - }The open forum is an opportunity to gather and discuss ideas, solutions and already existing ATs. Multiple participants of all stakeholder groups complimented the open forum. PWDs noted that finding solutions to their challenges independently is a challenging and tedious task. DO-1 described how the meeting helped: "They sat down together and discussed these [ATs], [...] and perhaps they have never even thought about it themselves, but then they see, aha, there is something, maybe I could use that for myself or for my clients." Moreover, it helps to enhance the initiative’s public outreach and popularity (PWD-2, PAT-2). It can also reduce barriers for other people with disabilities to share their challenges and participate in one of the lectures, a point highlighted as significant by PWD-1, PWD-2, and PWD-4.

\subsection{Influence on Ableism}

\textbf{Lectures create awareness - }Ten people of all stakeholder groups named the initiative's positive impact on creating awareness and broadening participants' horizons and knowledge about disability and ableism in society. Five interviewees said that disability is a topic of taboo, and the majority thought that a reason for this is because people have fears and do not know how to interact with people with disabilities. PAT-1 believes, "For many, this [lecture] is occasionally the first point of contact with people with disabilities or chosen forms of disabilities." ST-3 said that talking with people with disabilities is crucial "because I believe that only when you hear it [...], then you get access to it." PWD-2 and PWD-4 said, they often feel treated differently and considered less able, especially regarding knowledge and skills, but experienced the opposite during the course. PWD-2 said, "I believe it's because the framework of the course is chosen in a way that there are no reservations or apprehensions, as it takes place on a professional level." Six interviewees (PAT, STs, PWDs, DO) believe that participation in the lecture will have a lasting experience on participants, making them more sensible in the future. 

\textbf{Initiative does not reach enough people - }While PWD-4 thought, "If you can just get one person to reconsider, then it's worth it," others felt it will not change albeist views in general. PWD-4 and ST-4 believe that only those students who are already open-minded and have a certain awareness about the topic will participate in the Master course. This emphasizes the importance of working with people with disabilities in mandatory courses, like the Bachelor’s course. Three PWDs expressed doubt that the current size of the initiative is sufficient to achieve a meaningful change. It was them who also stressed the importance of networking and public outreach to potentially reach individuals who are not actively participating in the lectures. The DO-1 has the hope that "The more frequently one is confronted with these issues, the more it becomes natural, and at times, one might find themselves automatically doing things right." 

\textbf{ATs do not mitigate ableism - }From the above as well as opinions of multiple interviewees it needs to be stressed that access to appropriate ATs alone does not improve the situation of people with disabilities (PAT-1, PA-1, PWD-1, ST-4, ST-5). As some said, ATs can facilitate participation for people with disabilities, but as explicitly stated by PA-1 "[…] achieving inclusion through technologization is simply an illusion". When organizing co-design workshops of any kind, it is therefore equally important to emphasize joint learning, exchange, and discussion.

\section{Discussion}

The presented initiative strives to develop appropriate ATs by co-designing them in a university setting with students and people with disabilities. Interviews with participants showed that developing fully functional and market-ready ATs is difficult to achieve within the given time of a lecture. To make solutions usable and fitting to requirements, developments are continued with other students or the student technicians. Similar as \cite{neil_squire_society_makers_2024}, the initiative will share solutions online with a list of materials and instructions for use and building. Together with associations and AT professionals, we will investigate how to possibly adapt solutions within the structure of the initiative. The interviews imply that the initiative's co-design approach fostered a safe and anti-ableist environment. However, since only a small fraction of the 70 participants were interviewed, we will consistently encourage participants to provide personal or anonymous feedback.


One often-named limitation by interviewees is that some people with disabilities might not be able or do not want to physically participate in a lecture. Although we believe that personal meetings and exchanges are core elements of the initiative, we will explore a combination of in-person and virtual meetings in the future. To extends its application and number of participants, we plan to integrate co-design projects into courses across various disciplines while keeping the co-design procedure as explained above. The diverse range of ATs makes it feasible to identify projects suitable to the learning objectives of other courses.


All interviewees felt that the participation positively impacted views and attitudes towards people with disabilities. However, PWDs stated that, ableism can only be diminished if the political situation changes. This includes, for example, the right to fair working conditions, a better health care system, and support for education until university degrees. It, therefore, remains in doubt to which extent the initiative mitigates ableism in general. However, aligned with statements from interviewees, we believe that every step in the right direction does make a difference. It also stresses the importance of not only creating hackathons or university lectures for the purpose of developing ATs. They should rather be seen as platforms to bring awareness and foster exchange.



\section{Conclusion}

The presented initiative aims at developing ATs by uniting the regional community of people with disabilities, associations, AT researchers, and students during co-design lectures. Interviews with thirteen participants identified its advantages and challenges. While the development of ATs was successful regarding the incorporation of individual requirements, the time of the lectures, the grading constraints, and the background of the group members influenced design outcomes. It is, therefore, necessary to continue unfinished AT solutions beyond one semester. The interviews suggest that the initiative will positively impact views and attitudes towards people with disabilities for those actively participating. However, influence on other important stakeholders, like governments and health insurance, is limited. Therefore, we advocate for making the topic of disability an integral part of university education, emphasizing the importance of allowing people with disabilities to share their own experiences while continuing public outreach activities. 

\begin{acks}
    The authors would like to thank the University of Innsbruck, its Rectorate Team, and its partners for supporting the initiative and the lectures. This work and the initiative has been supported by the \grantsponsor{UIBK1669}{Universität Innsbruck - Förderkreis 1669}{https://www.uibk.ac.at/de/foerderkreis1669/} (\grantnum{UIBK1669}{2023-TECH-3}).
\end{acks}

\printbibliography



\end{document}